\documentclass[acmsmall]{acmart}

\def\BibTeX{{\rm B\kern-.05em{\sc i\kern-.025em b}\kern-.08emT\kern-.1667em\lower.7ex\hbox{E}\kern-.125emX}}
    
\copyrightyear{2018}
\acmYear{2018}
\setcopyright{acmlicensed}
\acmConference[DIS'19]{}{June 23--28, 2016}{San Jose, CA, USA}
\acmDOI{http://dx.doi.org/10.475/123_4}
\acmISBN{123-4567-24-567/08/06}

\begin{document}

%
% The "title" command has an optional parameter, allowing the author to define a "short title" to be used in page headers.
\title{Affordance Analysis of Virtual and Augmented Reality Mediated Communication}

%
% The "author" command and its associated commands are used to define the authors and their affiliations.
% Of note is the shared affiliation of the first two authors, and the "authornote" and "authornotemark" commands
% used to denote shared contribution to the research.
% \author{Mohammad Keshavarzi}
 %\email{mkeshavarzi@berkeley.edu}
% \affiliation{%

\author{Mohammad Keshavarzi}
\affiliation{
  \institution{University of California, Berkeley}
 % \streetaddress{1 Th{\o}rv{\"a}ld Circle}
 % \city{Hekla}
  \country{USA}}
  \email{mkeshavarzi@berkeley.edu}

\author{Michael Wu}
\affiliation{
  \institution{University of California, Berkeley}
 % \streetaddress{1 Th{\o}rv{\"a}ld Circle}
 % \city{Hekla}
  \country{USA}}
  \email{wuxiaohua1011@berkeley.edu}
  
  \author{Michael N. Chin}
\affiliation{
  \institution{University of California, Irvine}
 % \streetaddress{1 Th{\o}rv{\"a}ld Circle}
 % \city{Hekla}
  \country{USA}}
  \email{chinmn@uci.edu}
  
  \author{Robert N. Chin}
\affiliation{
  \institution{University of California, Santa Barbara}
 % \streetaddress{1 Th{\o}rv{\"a}ld Circle}
 % \city{Hekla}
  \country{USA}}
  \email{rnchin@engineering.ucsb.edu}
  
  \author{Allen Y. Yang}
\affiliation{
  \institution{University of California, Berkeley}
 % \streetaddress{1 Th{\o}rv{\"a}ld Circle}
 % \city{Hekla}
  \country{USA}}
  \email{yang@eecs.berkeley.edu}

%
% By default, the full list of authors will be used in the page headers. Often, this list is too long, and will overlap
% other information printed in the page headers. This command allows the author to define a more concise list
% of authors' names for this purpose.

%
% The abstract is a short summary of the work to be presented in the article.
\begin{abstract}
Virtual and augmented reality communication platforms are seen as promising modalities for next-generation remote face-to-face interactions. Our study attempts to explore non-verbal communication features in relation to their conversation context for virtual and augmented reality mediated communication settings. We perform a series of user experiments, triggering nine conversation tasks in 4 settings, each containing corresponding non-verbal communication features. Our results indicate that conversation types which involve less emotional engagement are more likely to be acceptable in virtual reality and augmented reality settings with low-fidelity avatar representation, compared to scenarios that involve high emotional engagement or intellectually difficult discussions. We further systematically analyze and rank the impact of low-fidelity representation of micro-expression, body scale, head pose, and hand gesture in affecting the user experience in one-on-one conversations, and validate that preserving micro-expression cues plays the most effective role in improving bi-directional conversations in future virtual and augmented reality settings.
\end{abstract}



%
% Keywords. The author(s) should pick words that accurately describe the work being
% presented. Separate the keywords with commas.

%
% This command processes the author and affiliation and title information and builds
% the first part of the formatted document.
\maketitle

\section{Introduction}
Human face-to-face interaction has played a critical role in allowing various sectors of the society to connect and interact effectively. With the development of modern \emph{computer mediated communication} (CMC) techniques, remote communication or telepresence has enabled users to interact over long distances and convey more non-verbal and emotional cues than conventional text or audio-only systems. So far, video conferencing from commercial providers such as Skype and Zoom owns a commanding share of all the possible CMC modalities. More recently, \emph{virtual reality} (VR) and \emph{augmented reality} (AR) systems have been introduced as new commercially viable communication platforms \cite{Beck2013, Billinghurst2002, Davis2014, Haxby2000}. Such modalities, often referred to as immersive telepresence, utilize head mounted displays (HMD) and body tracking technologies to reconstruct and represent head and body pose \cite{Garau2003,Johnson2015}, allowing users to remotely communicate using virtual avatars\cite{Alldieck_Xu_Theobalt_Pons-moll, Maimone_Fuchs_2011} or fully embodied 3D scans \cite{Blanche2010, haxby2002, Kraut2003}. The captured 3D information can go beyond the participants themselves and can include their surrounding environment, allowing higher levels of human engagement through spatial context \cite{Hormann1981}.

To this end, if available technologies could allow accurate reconstruction and representation of human interaction and further their surrounding environment, one may argue that immersive telepresence should be a valid surrogate to physical face-to-face interaction with full-fledged features to communicate both verbal and non-verbal cues. Nevertheless, the current limitations of the hardware and software in capturing, processing, and transferring such data still prohibit consumer-grade VR/AR systems to achieve high-enough fidelity 3D reconstruction and rendering. Therefore, while developing new VR/AR technologies to fill the critical gaps, it is important to follow user-centric design principles and study which communication features play a more effective role in virtual face-to-face communication, in order to allocate appropriate resources in various stages of the future telepresence development process.

Furthermore, we believe it is important to perform evaluation of available CMC platforms by considering their conversation context and discussion scenarios. During previous generations of CMC paradigm shifts, from audio-only to video-enabled platforms, experimental studies \cite{Kiyokawa2002, Maimone2012} showed that simply enabling remote collaboration through audio-visual platforms instead of audio-only methods does not significantly improve collaborative performance, especially not to the levels reported in co-present communications \cite{Gergle2004, Kohli2008}. Other studies have reported different performance rates for various conversation tasks, when comparing an audio-only medium to a video-enabled setting in workplace settings \cite{Clark1991}. 

As various conversation types require different levels of emotional and intellectual  engagement, investigating how communication cues facilitate in the productivity and performance of such conversation types is critical in evaluating the CMC platforms. The importance of qualitative non-verbal cues in CMCs' such as facial expressions\cite{Fish1992, Fuchs_State_Bazin_2014}, body pose and gestures \cite{Fussell2004} have been widely investigated in social and interactive contexts. Such cues have shown to inform individuals about the current states of the participants to an extent that self-awareness has been reported to be higher in CMC methods when compared to general face-to-face situations \cite{Kendon1967, Kirk2007}. Clark and Brennan \cite{Clark1996} discussed various non-verbal methods and their correlation with communication grounding (i.e. establishing mutual knowledge, belief, attitudes and expectations \cite{Bogo2015}) highlighting the role of symbolic gestures and facial expressions in grounding deictic references.

Our study attempts to explore non-verbal CMC features in relation to their conversation context. We investigate the correlation between conversation types and communication features by experimenting with virtual and augmented reality methods in addition to conventional 2D video conference settings for one-on-one meetings. Our goal is to measure the level of impact of various communication features in one-on-one meeting scenarios. Such understanding would allow researchers to effectively invest limited resources in developing and prioritize the most important communication qualities. We evaluate the importance of these communication features in both participant\textquotesingle s subjective impressions and a detailed analysis of their preferences in actual verbal and nonverbal communication tasks. 

\subsection{Related Work}
Previous work related to this study can be explored in two main categories. First, understanding how various non-verbal communication cues affect conversation quality and task performance in general CMC use cases, and second, exploring VR and AR based communication platforms, specifically, to evaluate how certain technological features enhance communication performance in such settings. A majority of the literature related to non-verbal features in CMC's lies in investigating how video enabled settings enhance different communication scenarios compared to audio-only platforms. As video feeds capture a much wider array of nonverbal behavior and transfer critical communication cues such as body gestures, facial expressions and eye gaze, various experiments \cite{Gergle2004}\cite{doherty1997face} have concluded that having full visual representation of the other party can improve task performance and support communication grounding %MK:Grounding has been explained in the intro paragraphs% 
of participants. Moreover, exploring the role of the body gestures in CMC's has also been widely investigated in multiple studies.\cite{Streeck1993, Fussell2004} A study by Beck et. al found that tracing gesture and pointing is independent of whether two subjects are remote or local\cite{Beck_Kunert_Kulik_Froehlich_2013}. Clark and Krych performed experiments of helper(the expert) and worker based scenarios, reporting that through the aid of video feeds and gestures, partners were likely to identify and resolve related problems more efficiently, compared to audio-only communication settings. \cite{Clark2004}. While our work also studies body gestures as a contributing communication factor, we focus on face-to-face meeting scenarios which involve both parties to convey information in a rather balanced conversation context compared to a helper-worker scenario which one party has more outward expressions to the other. 

%Fullwood and Doherty-Sneddon \cite{Fullwood2006} studied the effect of capturing eye-gaze over a video link . Their experiments showed that lack of gaze during a conversation has a negative impact on subjects' information recall.

Furthermore, presence of facial expressions has been found to be a important communication factor within CMC conversation scenarios. In their study of user embodiment through avatars in collaborative networked VR environments, Benford et. al\cite{Benford:1997:EAC:959145.959151} emphasized that facial expressions were the primary means of expressing or reading emotion from a conversation partner, being something of a more refined complement to gestures. Much of the importance of facial expressions was derived from the ability to reflect not just what the user explicitly controlled, but also subconscious and involuntary movements that were equally critical to mutual understanding during collaborative tasks. They found that in virtual avatars, it was especially difficult to reflect these involuntary expressions without a engineering a complicated system capable of recording and rendering the user's face onto their corresponding avatar in real time.

Evaluating collaborative communication through virtual and augmented reality platforms have been mostly explored through the lens of innovating new technological features\cite{Fuchs2014}\cite{Beck_Kunert_Kulik_Froehlich_2013}. While VR and AR hardware have dramatically evolved in the past few years, from large CAVE rooms, to lightweight head mounted displays, the affordance of each of these media is highly dependent to the level of communication factors that can be transferred through these environments\cite{short1976social}. Kiyokawa et. al \cite{Kiyokawa2002} conducted a task performance study that compared various AR configurations in head mounted devices. They found the more
difficult it was to use non-verbal communication cues due to hardware constrains, the more people resorted to speech cues to compensate. While our study also experiments different combinations of communication features, we rely on user preference than measuring specific task performance.

Moreover, studies of comparing AR systems and desktop systems have also been conducted in multiple user experiments. Billinghurst et. al \cite{Billinghurst2002} developed a tangible augmented reality system for face-to-face interaction and report that in a collaborative AR environment subjects tend to behave more similar to an unmediated face-to-face meeting compared to 2D screen CMC interfaces. Kato and et. al \cite{article} studied an AR video conferencing system where AR users can augment remote participants on paper cards while desktop users can understand the conversation relationship by observing AR users' face angle. They concluded that even if head-mounted devices could not track eye movements for the AR users, video images still could play an important role for both AR and desktop users. Piumsomboon\cite{Piumsomboon2018} used a system called Mini-Me, which had one participant in AR and the other in VR and conducted a user study comparing collaboration in both asymmetric and symmetric conditions. Their study suggests that social presence and the overall experience of mixed reality collaboration in either condition could be significantly improved when gaze and gestures were available. In our study, in addition to investigating the relationship between non-verbal cues and virtual configurations, we explore how and to what degree each non-verbal communication feature contributes to various conversation scenarios

In VR communication settings, as participants are rendered through virtual avatars, various studies have explored different properties and non-verbal features of such avatars and their impact in virtual communication. Smith and Neff \cite{Smith2018} experimented embodied virtual reality settings where subjects' movements are rendered onto an avatar using motion capture suits, thus supporting body language as nonverbal behavior alongside with verbal communication. They conclude that the conversation pattern in embodied avatars is more similar to face-to-face interaction as it provides a higher level of social presence. While in contrast, providing only the environment without embodied virtual reality generally lead to the feeling of loneliness and degraded communication experience. Garau et al. compared the effect of a humanoid avatar with basic, genderless visual appearance to an avatar with more photorealistic and gender-specific appearance. Also, they examined the difference between random eye gaze to inferred eye gaze from voice. They discussed the importance of aligning visual and behavioral realism for increased avatar effectiveness and collected and analyzed data that suggest that inferred gaze significantly outperform random gaze\cite{Garau2003}. Slater et. al  study group behavior in VR and concluded that even with a simple avatar, socially conditioned responses such as embarrassment can be generated; they also found a positive relationship between the presence of being in place and co-presence\cite{Slater2000}.

\begin{figure}
  \centering
  \includegraphics[width=\columnwidth]{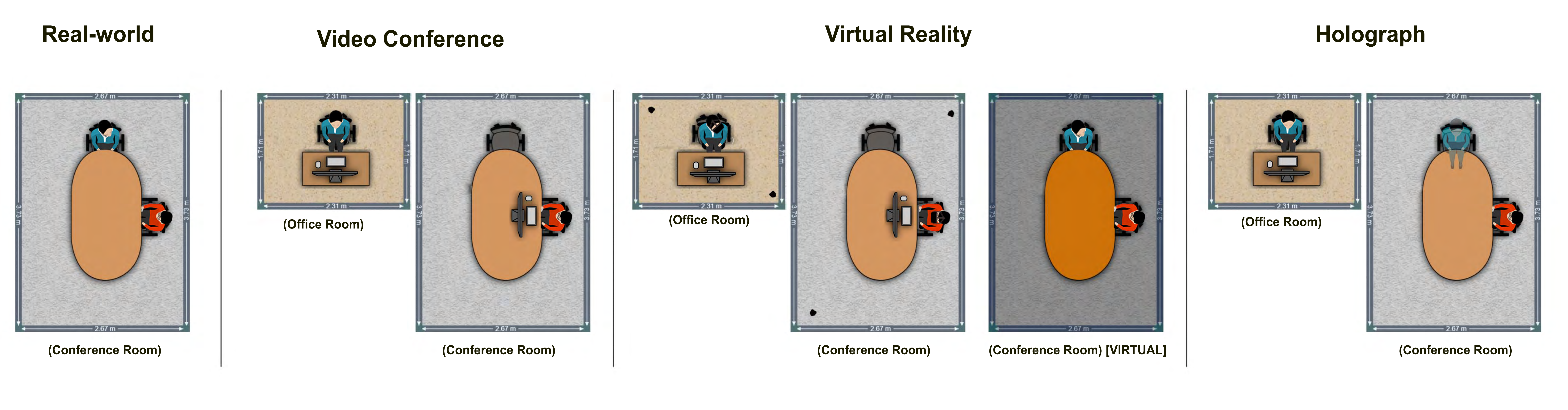}
  \caption{Overhead diagrams showing the room layout and positions of the interviewer (in blue) and interviewee (in red) for each of the four experimental settings: real-world, video conference (2D video), virtual reality, and holograph.}\label{fig:figure_setting_top_down}
\end{figure}
\section{Methods}

\begin{figure}
    \centering
   \includegraphics[width=\textwidth]{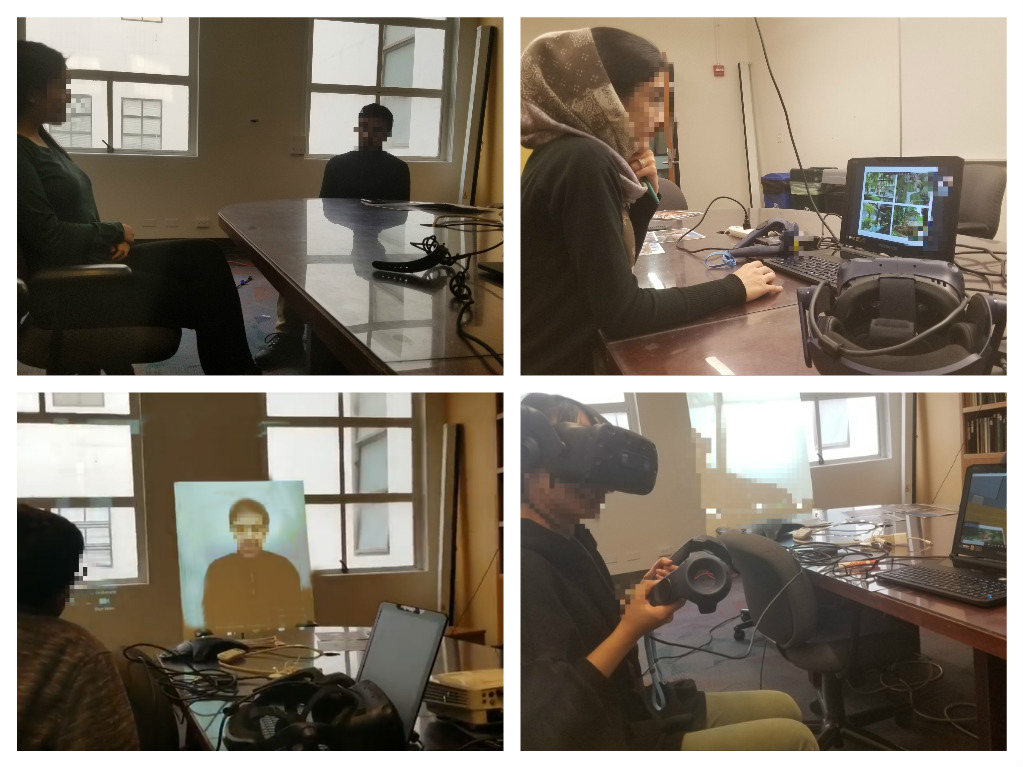}
   \caption{Images of participants in each of the four settings. Top left: Real-world interview. Top right: 2D video interview. Bottom left: Hologram interview. Bottom right: Virtual reality interview.} \label{fig:figure_actual_setting}
  \end{figure}

We design a study to evaluate virtual forms of communications and their corresponding features against various one-on-one meeting conversation scenarios. Four settings are prepared as follows: real world setting as a gold standard, a conventional video conference setting using personal computers with 2D screens(ex. Skype, Zoom), a virtual reality setting, and a holographic projection setting (see Figure ~\ref{fig:figure_setting_top_down} and Figure ~\ref{fig:figure_actual_setting}). We examine how the presence of communication features such as micro-expressions, 1:1 body scale, head pose and body gestures affect the user experience within different conversation types.

\begin{figure}
\centering
  \includegraphics[width=\columnwidth]{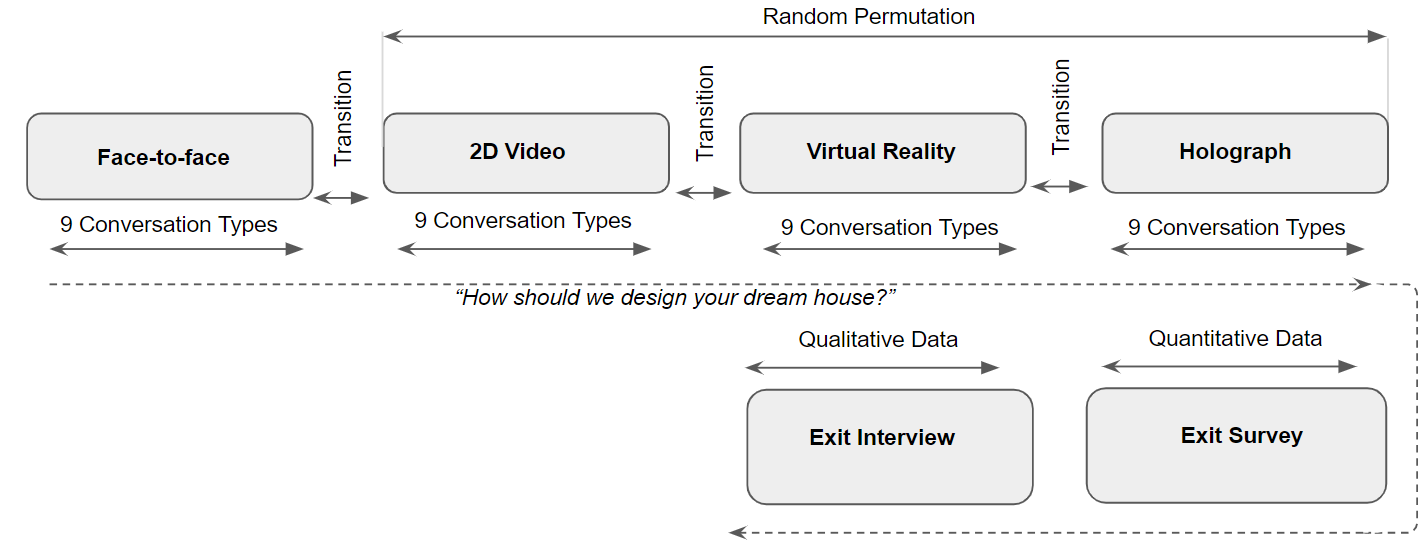}
  \caption{Linear flowchart demonstrating the experiment workflow starting with conversation in through Face-to-Face, 2D video, Virtual Reality, and holograph settings. Ending with a Survey and Exit Interview.}\label{fig:figure_linear_flowchart}
\end{figure}

\subsection{Conversation Types}
\label{sec:conversation-type}
To address different conversation types, we use the criteria presented in Clark's study \cite{Clark1991} for workplace related scenarios. To trigger all target conversation types during our experiment, we develop an architect-client scenario in which we hold a 60 minute interview with the subject on the topic: "How would you want your dream house to be designed?" Utilizing this context to integrate various design and decision making tasks, we trigger all of the objective conversation types over the course of the interview.  Table \ref{tab:table_intro} describes how each conversation type was prompted during the architect-client scenario.

The interview is divided into four consecutive 15-minute segments, one corresponding to each setting.(see Figure ~\ref{fig:figure_linear_flowchart} The real world setting is always presented first, and the ordering of the remaining three settings is randomized. After each segment, we transition smoothly and rapidly to the next setting, the conversation continuing from where it left off in the previous setting. The real location of the subject remains constant throughout all four of the settings(see Figure ~\ref{fig:figure_setting_top_down}), while the interviewer is displaced to a remote location for communication through the three virtual settings. 

\subsubsection{Participants}
Our recruitment screening required participants to speak fluent English, not have a history of photosensitive seizures, and not suffer from any medical conditions affecting cardiovascular, pulmonary, nervous or musculoskeletal systems or other physical conditions which may lead to danger or discomfort in virtual reality environments. A total of 9 subjects (4 male, 5 female) were recruited through on-campus advertisement, email lists and social media postings Participants were aged 19-30 ($\mu$ = 22.56, $\sigma$ = 4.07). Each interview for this experiment took approximately 1 hour to complete and IRB approval was maintained ahead of the experiment. Participants were compensated with a free meal after their interview. All participants successfully completed the interview without any unexpected terminations.

\begin{table}
      \caption{Triggering methods for each conversation type. All conversation types were triggered at least once in each communication setting.} \label{tab:table_intro}
\begingroup
\setlength{\tabcolsep}{10pt} % Default value: 6pt
\renewcommand{\arraystretch}{1.3}
    \begin{tabular}{p{5cm}  p{7cm} }
    \hline
    {\small\textbf{Conversation Type}} & {\small\textbf{Description}} \\ \hline
 {\small{Making decisions}} & {\small{Subject is shown different pictures and/or design proposals. After a few minutes of discussion, it is asked to decide between options}} \\
 
  {\small{Generating ideas}} & {\small{Subject is asked to propose layout ideas for the discussed space (e.g. Living room, bedroom etc.)}} \\
  
  {\small{Resolving disagreements}} & {\small{Although the interviewer is aware of the subjects' preference in the design option, s/he propose an opposite design proposal to generate disagreement and }} \\
  
  {\small{Exchange personal information}} & {\small{Questions about personal habits and social preferences are asked}} \\
  
  {\small{Checking project status}} & {\small{Interviewer asks questions about how the tasks are proceeding in various settings}} \\
  
   {\small{Negotiation}} & {\small{Disagreements are triggered intentionally by the interviewer to provide a negotiation/bargaining discussion context}} \\
   
    {\small{Asking questions}} & {\small{General questions about design and color preferences are asked}} \\
    
   {\small{Explain a difficult concept}} & {\small{Subjects are asked to explain their personal level of modernism or classicism in addition to subject such as sustainability and privacy for the house layout planning}} \\
   \hline
    \end{tabular}
    \endgroup
    
\end{table}

\subsection{Communication Settings}

\subsubsection{Real-world setting}

We perform the real-world setting interview in a standard conference room, comprising of a large table and two chairs. After being prompted to enter the room and being greeted by the interviewer, the subject is seated on a chair positioned to maintain a constant perceived distance to the interviewer in all following virtual settings. The interviewer and interviewee remain seated when undergoing conversation.  No digital tools are used for communication in the real-world setting; printed photos, a pen,  and paper serve as the only non-verbal media of communication between the subject and interviewer. Subjects can take notes of the conversation and use a pen to annotate illustrations during the collaborative design tasks.

\subsubsection{2D video conference setting}
A conventional video conferencing software - in our case, Zoom - is used to establish virtual communication between the subject and the interviewer. The subject maintains their location in the conference room, while the interviewer is located in another room. Each party uses an identical 15" laptop and front facing webcam to continue the interview in a 2D video conference setting. The size and scale of the interviewer does not reflect the actual dimensions perceived during the in-person segment of the interview. For collaborative design tasks, the screen of interviewer is shared and displayed on the subject's laptop screen. Using native annotation tools provided by the software, the subject is able to draw, annotate or label information on the computer screen.

  \begin{table}
          \captionof{table}{Availability of each non-verbal communication feature in experiment settings.} \label{tab:settings_available_in_settings}
    \begin{tabular}{l r r r}
        % \toprule
          & & \multicolumn{2}{c}{\small{\textbf{Medium}}} \\
       \cmidrule(r){2-4}
        {\small\textit{Communication Feature}}
        & {\small \textit{Video}}
          & {\small \textit{Holograph}}
        & {\small \textit{VR}} \\
        \midrule
        Micro-expressions & Yes & Yes & No \\
        Body Scale & No & Yes & Yes \\
        Head pose & No & No & Yes \\
        Body gesture & No & Yes & Yes
        % \bottomrule
      \end{tabular}
  \end{table}

\subsubsection{Virtual Reality setting}
Our virtual reality setting uses a computer-generated virtual conference room designed to mimic the layout of the real-world setting in look, feel, and scale. Both the subject and interviewer use Vive Pro virtual reality headsets and pairs of handheld controllers tof  control virtual avatars that reflect their bodily movements. Verbal communication is enabled with VoIP, utilizing the Vive Pro headset integrated microphone and headphones. The seating locations of the interviewer and subject reflect the real-world settings. For the avatars, head pose is calculated using the headset tracking system, and body gestures are estimated with inverse kinematics methods that read positional input from the handheld controllers.

\subsubsection{Holographic setting}
Ideally, a 3D holographic avatar of a human can be visualized on a quality wearable AR device or a holographic display system. However, the challenge is to capture those 3D images in real time. Typically, a fairly expensive, experimental camera stage paired with high-speed computers are needed to achieve the performance for our study in this paper. As a trade-off, instead, we design and implement a 1-to-1 scale 2D projection representation for the remote participant, making it appear as if the interviewer is seated at the table. The projection is calibrated to make the captured video data from the remote interviewer reflect their real-world scale. The seating position of the interviewer and subject mimics the real-world setting. A speaker is placed in front of the projection to mimic the spatial sound settings of the participants.  For collaborative design tasks, the subject uses a whiteboard and marker pen for annotations. We believe in the scenario of one-on-one sit-down interview scenario, the 2D projection method is a good surrogate to the more advanced 3D hologram in AR.

\begin{figure}
  \centering
  \includegraphics[width=\columnwidth]{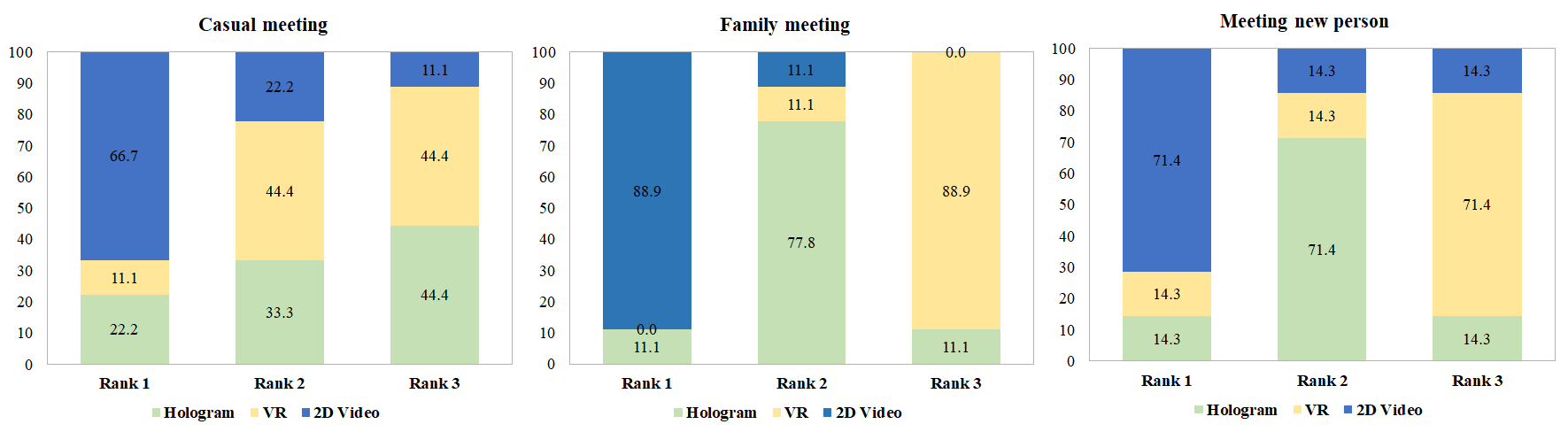}
  \caption{Stacked bar charts showing the rankings for each of the three settings (2D video, virtual reality, hologram) based on three questions asking participants which settings they would choose as their first, second, and third choice for various types of meetings. Left: Question about a casual meeting. Center: Question about a family meeting. Right: Question about meeting an new person.}\label{fig:stacked_bar}
\end{figure}
\section{Results}

We analyze our findings using the compiled sets of quantitative direct and indirect questions, in addition to discussing descriptive feedback that was received after the questionnaire. The different layers of collected data (direct questionnaires, indirect analysis, and descriptive exit interviews) are compared and contrasted to see how well they align with one another. Three sets of anonymous surveys are performed at the end of the experiment. First, participants are asked to rank their preferred methods of communication outside of real life face-to-face conversation for three different scenarios. These meeting types (casual meeting, family meeting, and meeting a new person) reflect different subsegments of the conversation tasks criteria, allowing a broad understanding of the CMC affordances in a contextual fashion. Second, participants are asked to define the level of importance of four communication characteristics- gestures, head pose, head/body scale, and micro-expressions- in a five point Likert scale.  (See Table \ref{tab:settings_available_in_settings}) is clearly communicated to the participants in the questionnaire. Finally, subjects are asked to define whether they prefer to perform specific conversation tasks in each CMC settings. In this segment, a binary questionnaire is conducted (Figure \ref{fig:stacked_bar}) and subjects used Yes or No statement to indicate their readiness of such tasks. All conversation types are triggered during the interview and no direct explanation is given to participants to refer to any segment of the interview itself.

\subsection{Communication Types}

Our analysis on how well each CMC performs in relation to a conversation context is performed in two separate comparative and non-comparative studies. In the first study, participants are asked to compare and rank the CMC media of hologram, virtual reality, and 2D video as their first, second and last choice in general conversation contexts. Furthermore, independent to other CMC media, subjects were asked to specifically define their willingness to execute more detailed conversation criteria in a binary fashion.
\begin{figure}
  \centering
  \includegraphics[width=\columnwidth]{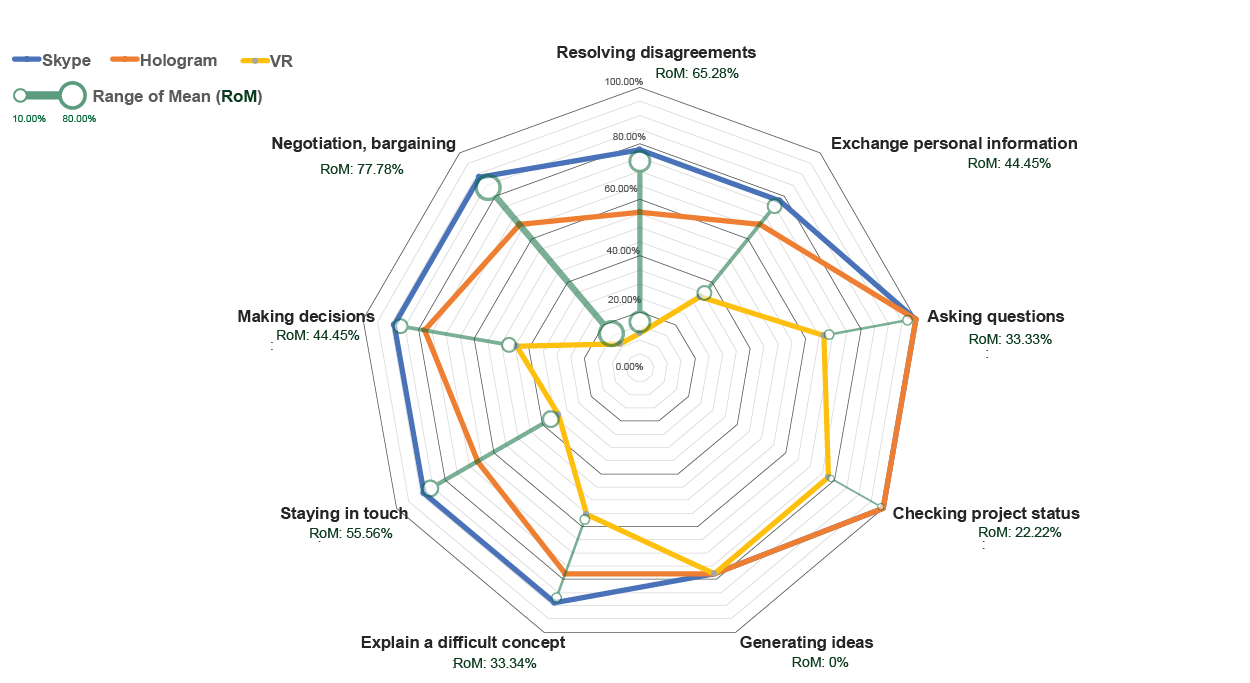}
  \caption{Acceptance rate of three communication settings (2D video, hologram, virtual reality) in various conversation contexts. Participants were asked whether they are willing to use the mentioned CMCs in nine conversation scenarios.} \label{fig:fig_big_spider}
\end{figure}

\subsubsection{Comparative analysis}
When comparing CMC platforms for a casual meeting scenario, as illustrated in Figure \ref{fig:stacked_bar} the distribution of the ranking of the hologram method by respondents has the greatest variance at Var(X) = 0.62. The quartile distributions show that over 75\% of participants rank 2D video as their number one choice for a casual meeting. On the other hand participants narrowly prefer hologram over virtual reality for casual meetings with a 0.11 difference in the mean rank score. Less than 25\% of participants choose virtual reality or hologram as their number one choice for a casual meeting. For family meetings, when excluding the outliers, 100\% of respondents rank 2D video as their first choice, hologram as their second choice, and virtual reality as their last choice. Similar results are observed When asked about meeting a new person, over 70\% of respondents rank 2D video as their first choice ranked virtual reality as their last choice.  
\begin{table}
  \caption{Compiled participant acceptance rates for conversation types (emotional vs non-emotional) tested in each experimental medium (2D video, holograph, virtual reality).}\label{tab:table_approval}
  \centering
  \begin{tabular}{l r r r}
    % \toprule
      & & \multicolumn{2}{c}{\small{\textbf{Medium}}} \\
   \cmidrule(r){3-4}
    {\small\textit{Conversation Type}}
    & {\small \textit{2D video}}
      & {\small \textit{Holograph}}
    & {\small \textit{Virtual Reality}} \\
    \midrule
    Emotional Only & 86.11\% & 66.67\% & 25.35\% \\
    Non-Emotional Only & 88.89\% & 84.45\% & 62.22\% \\
    Overall & 87.66\% & 76.55\% & 45.83\% \\
    Emo./Non-Emo. Diff. & 2.78\% & 17.78\% & 36.87\%
    % \bottomrule
  \end{tabular}
\end{table}

\subsubsection{Context aware analysis} Moreover, we narrow down the conversation contexts into specific conversation types explained in Section \ref{sec:conversation-type}. By conducting a binary questionnaire we observe the correlation between communication media and conversation types. As seen in Figure \ref{fig:feature_range}, while 2D video has the highest mean acceptance rate in all conversation types, the difference between CMC modalities significantly decrease in conversation types which involve less emotional engagement between the two parties. In addition, the acceptance rate of Hologram and VR increases in less emotionally engaged conversation tasks, indicating that the current implementation of such CMCs' can be practiced in these conversation settings. 

\subsection{Communication Features}

We further analyze the impact of four non-verbal communication features (head pose, facial micro-expression, gesture, body scale) and evaluate their roles in maintaining CMC acceptability in various conversation types. Unlike communication settings in which we perform a comparative analysis by directly asking participants to rank their preference, we ask subjects to define the importance of each setting and then validate their answers based on an indirect analysis of their response of CMC settings (each containing certain communication features) in relation to the conversation types. In the initial survey, users respond to a Likert scale questionnaire indicating the level of importance of each communication feature based on the interview experience. As part of the survey, subjects are reminded whether the communication feature is present in each CMC or not.

\begin{figure}
  \centering
  \includegraphics[width=1.2\columnwidth]{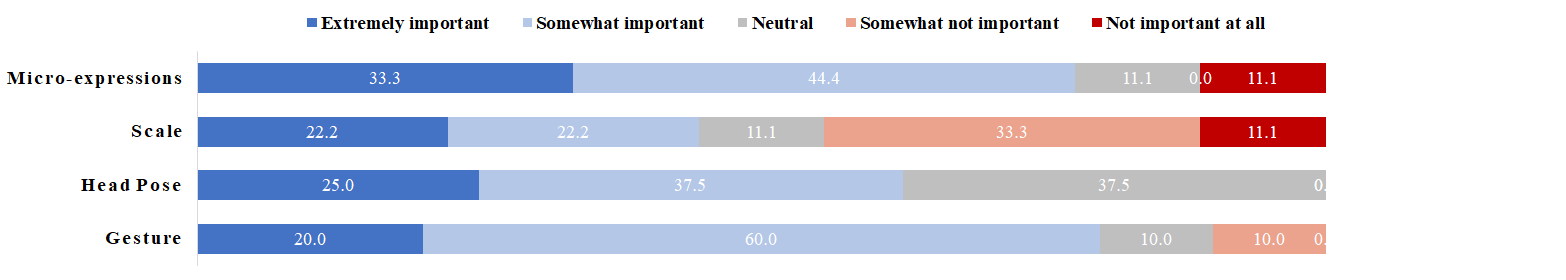}
  \caption{Likert scale of four different setting characteristics: gesture, head pose, body scale, and micro-expressions. Participants rated the importance of each feature in conversations from extremely important (solid blue) to not important at all (solid red).}\label{fig:likert_scale}
\end{figure}

\subsubsection{Direct questionnaires}
From Figure \ref{fig:likert_scale}, we observe the gestures category has the highest overall percentage of participants that indicated a degree of confidence in the importance of the characteristic, at 80\%. It is followed by micro-expressions with 77.78\%, and head/body scale with 62.5\% of participants indicating importance. The body scale characteristic is the only category in which the majority of participants did not indicate importance, with only 44.44\%.

\subsubsection{Indirect analysis} Although the availability of the each feature in the CMCs is communicated with the subjects, the ability to independently distinguish the impact of each communication feature is not possible due to the experiment workflow. As we do not modify general CMC characteristics in the experiment, participants do not experience each communication feature independently from the other features. Therefore, we perform an indirect analysis of binary questionnaire addressing CMCs in relation to their conversation context. Given the presence of each communication feature in the CMC, we can calculate the expected range of CMC features scores for each conversation type, and compare the findings with the direct questionnaires of the importance of the CMC features.  

\begin{equation}{\label{eq:minimization}}
\begin{aligned}
\min \quad & (M + G + S + Hp + U - 100)^{2} \\
\textrm{s.t.} \quad & Vi -  \varepsilon \leq M + U  \leq  Vi + \varepsilon \\
                    & Holo - \varepsilon\leq M + G + S + U  \leq Holo + \varepsilon\\
                    & VR - \varepsilon\leq G + S + Hp + U  \leq  VR + \varepsilon \\
\end{aligned}
\end{equation}

Equation \eqref{eq:minimization} assumes that the four features (namely, Micro-Expression ($M$), Gesture ($G$), Scale ($S$), and Head-pose ($Hp$)) contribute to different settings along with other unknown factors ($U$) that we do not explicitly consider to validate in different media (e.g., voice, pictures). Each variable is bounded between 0 and 100, and so are all the inequalities constraints.  The  objective is to minimize the deviation from our ``gold standard'', the face-to-face setting.  The data for the variable video ($Vi$), hologram ($Holo$), and VR ($VR$) is extracted from calculating the average of our subject's response of whether they will use a certain medium for a specific task. Furthermore, we included a variable named $\varepsilon $ to represent the uncertainty that our subjects may have brought, and we assumed the $\varepsilon$ to be 20\%.\footnote{The uncertainty of 20\% is conveniently chosen such that the above minimization problem \eqref{eq:minimization} would yield possible solutions. If $\varepsilon$ is not considered then the problem in our experiment had no solution with the average values of $Vi$, $Holo$, $VR$ from the limited subject population.}

\begin{figure}
  \centering
  \includegraphics[width=\columnwidth]{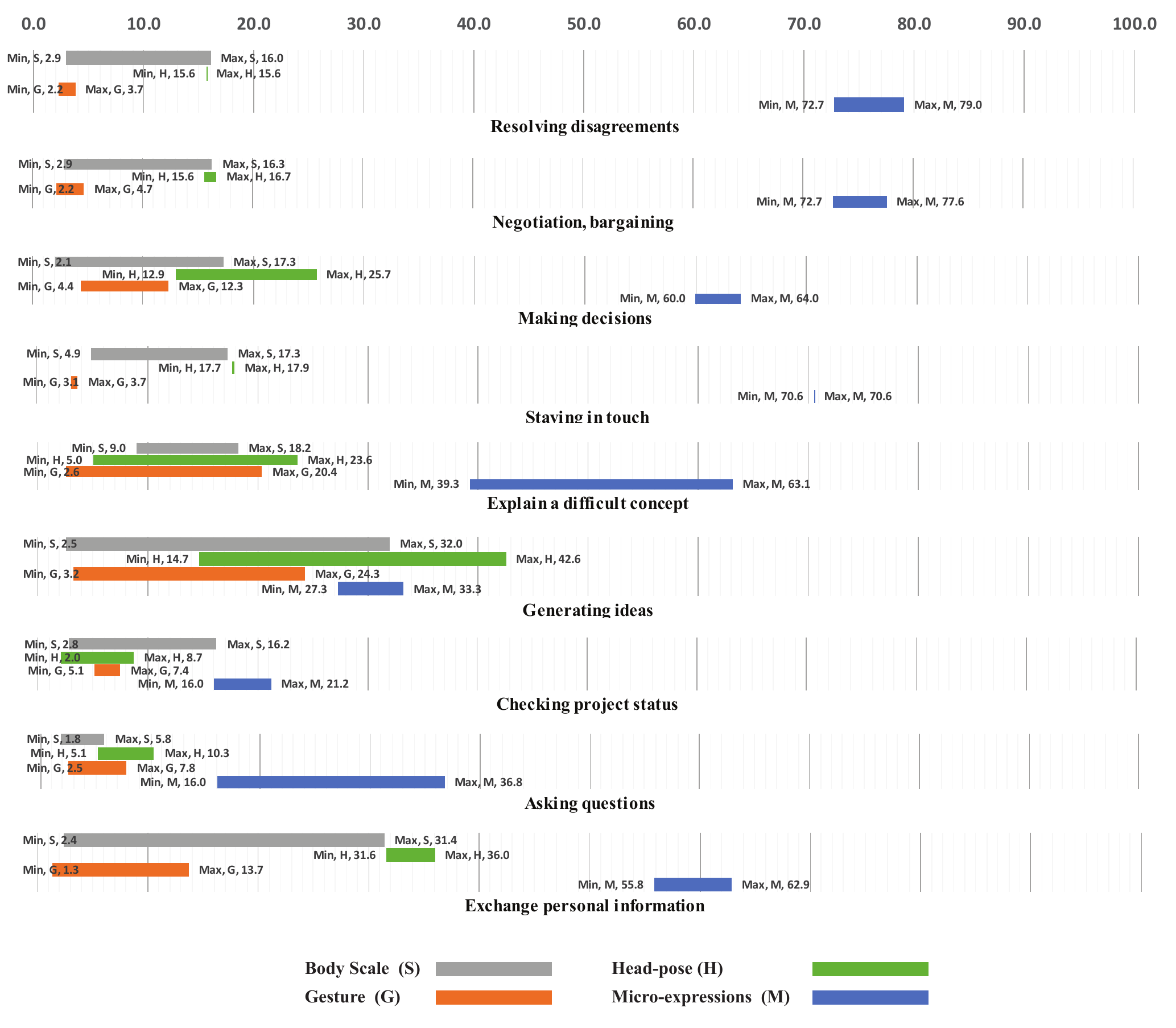}
  \caption{ Indirect analysis of communication features in various conversation types.}\label{fig:feature_range}
\end{figure}

The result, as portrayed in Figure \ref{fig:feature_range}, demonstrates that our subjects' conformity with regard to the relative importance of micro-expression is generally higher in tasks that involves emotion. Also, micro-expression attains the highest minimum value in all tasks, which implies that regardless of tasks, our subjects recognize micro-expression as a feature that they want the most in conferencing. Furthermore, the ranges of the other features are generally greater in non-emotional tasks, with exception of Asking Question, than the ranges in emotional tasks. This phenomenon implies that users aren't as particular about experiencing the detailed features in non-emotional tasks as they would for tasks that would involve emotion, which validates what our subjects have commented in the Exit Interview phase. 

\subsubsection{Micro-expressions}
%TODO insert figure reference here
The direct questionnaires and indirect analysis have closely correlating results on facial micro-expressions. We observe micro-expressions maintaining a larger impact especially in emotionally engaged and intellectually difficult scenarios. In addition, the overall participant approval rates are very low for settings in which micro expressions are absent compared to those that include micro-expressions. 

Such conclusions also reflect the subjects' comments. P5 comments: "[I would like to] at least know whether the person is smiling, frowning, bored, etc. [It] is pretty important during conversation." P2 expresses a similar thought: "The lack of micro-expressions also hinders the way conversion is normally constructed." Referring to scenarios in which they would use VR for their own meetings, P2 also prefer to be able to "read between the lines" to see "oh, maybe the client doesn't like this idea. But if you can't see the frowning or nodding, the little gestures might get lost". Although the avatar in VR is capable of simulating mouth movements based on speech, it is not capable of rendering the wider variety of detailed facial expressions that P5 has stated that they would have liked to see. The lack of representation of micro-expression specifically in the VR setting of our experiment may have had an impact on how people perceived the usability of VR for both casual and more formal conversation.

When rendering the avatars and environment for VR, we have opted to use flat shading techniques over physically based rendering methods available at the time of the experiment, and the avatar models are strictly limited to 20,000 polygons each. Consequently, the interviewer's avatar and the interviewing room obviously look computer-generated. P8 comments that "the settings in Virtual Reality, although mimicking the actual setting, looks rendered and does not feel real enough that I feel comfortable in doing a business meeting there." P2 also shares this opinion, stating that "animated figures are really hard for me to take seriously...it can convey messages, but for a formal conversation, it lacks certain seriousness or authority."

\subsubsection{Body gestures}

Due to the absence of significant correlation between direct questionnaires and indirect analysis of the data for this segment, we are unable to draw definitive conclusions when attempting to compare them. The VR implementation integrates inverse kinematics techniques for motion tracking of the head, limbs, and upper body. This may result in inaccurate simulations of the body gestures, making them appear very unnatural to the participants. In addition, some of the participants can be noticeably perplexed regarding what actions are physically possible within the VR environment. P6 references this confusion: "I didn't know whether the other person [could] clearly see what I am doing in the virtual setting. And I think these movements are important during a meeting." P4 forgets that objects on the table could be picked up using certain controller buttons, and had to be reminded in the experiment that how this action was executable. 
  
However, some participants also acclaims VR's potential to be beneficial for interactive tasks requiring the caption of movements(such as pointing) that involve the entire body. P7 states: "I like that I am able to lift that paper up in VR, and that I can point at things." Furthermore, P7 claims that he "might use VR for things that might not be so convenient during a 2D video meeting." These statements are corroborated by the survey data, where higher preference rates are seen in tasks involving a possible need for users to point at objects. Other subjects implied that they would use VR for tasks in which they want to focus mainly on what the other person is physically doing, as opposed to attempting to discern what they are thinking. This also reflects the findings of the survey data, which state that gestures are more useful for tasks that are heavy on outward expressions while not mandating critical understanding of the other party's inner feelings.

\subsubsection{Body scale}
A close correlation is observed between the approval rates of both the  direct questionnaires and indirect analysis for this segment. On its own, the indirect analysis demonstrates negligible differences in approval rates when comparing settings that visually reflect real-life head and body scale to those that do not. Accurate body scale is found to be the least important factor among all of the examined communication features, providing support for the argument that the perceived real-life head and body scale of the interviewer is not a critical factor for communication.

\subsubsection{Head pose}

While head pose can be grouped in with body gestures, we have observed that there are some distinguishing factors for head pose on its own. The head pose state is dependent on the participants' locations in a CMC setting, whereas body gestures are considered independent of the locations of each participant. In addition, when conducting the experiment in the holographic setting, we deliberately position the interviewer's head pose so that it would not be visually aligned with the interviewees. Upon entering the holographic setting in a portion of his interview, P5 immediately asks "so how do you see me?", indicating that he has observed the interviewer appearing to be not looking directly at him through the projection. Although it is not directly stated, the participants have implied that they feel that the presence of head pose (or lack thereof) has very little impact on their ability to communicate, regardless of whether the tasks are emotionally involved. This is corroborated by the results from the direct questionnaires. 
 
Non-emotional task acceptance rates largely remain both high and constant regardless of whether any single characteristic is present. As a result, it is more logical to compare only the mean acceptance values from the emotional tasks to the direct Likert scale participant responses from the last section. As shown in Figure ~\ref{fig:feature_range}, for head pose, head/body scale, and micro-expressions, the mean emotional task acceptance rates in the presence of a characteristic differ by less than 5\% overall approval when compared to the percentage of participants that have directly indicated that they believe the characteristic is at least somewhat important. Gestures are the outlier in this comparison, with the direct participant response indicating an almost 34\% greater approval rate than suggested by the data. In this case, the mean emotional task acceptance rate may have been driven down by participants specifically having bad experiences in VR, where limitations in 3D avatar limb articulation are wholly evident, so the comparison for gestures is inconclusive.

By comparing the data from the direct questionnaire and indirect analysis of conversation contexts, we can validate that micro-expressions have higher impact in providing an efficient simulated digital communication experience than any of the other characteristics tested. Unlike any of the other characteristics, overall mean acceptance rates in the absence of micro-expressions are always substantially lower than when they are present; this differential is much more pronounced when only comparing the acceptance rates for emotionally involved tasks. This also seems to align with how a significant majority of the direct participant responses to the final Likert scale section have indicated that micro-expressions are at least somewhat important.

\section{Conclusion}

Our study has explored current AR and VR mediated communication in relation to general conversation contexts. By performing a series of user experience studies, we trigger these nine conversation types in different CMC settings and investigate non-verbal communication features through direct questioners and indirect data analysis. Our results have indicated that conversation types that mandate critical thinking or involve more emotionally engaged discussions are preferred to be executed in a setting that allows participants to convey a large variety of facial and micro expressions. We also observe that increased ease of perceiving facial expressions from the other party is more conducive to successful interactions in emotionally involved conversations (including bidirectional interactions such negotiations and solving/resolving disagreements). We have observed preserving micro-expression cues plays a more effective role in maintaining bi-directional one-on-one conversations in future virtual and augmented reality settings, compared to other non-verbal qualitative features such as realistic body scale, head pose and body gestures.

%
% The next two lines define the bibliography style to be used, and the bibliography file.
\bibliographystyle{ACM-Reference-Format}
\bibliography{sample}

\end{document}